\documentclass[12pt]{article}
\usepackage[totalheight = 23cm, totalwidth = 17cm]{geometry}

\newcommand{\be}{\begin{equation}}
\newcommand{\ee}{\end{equation}}
\newcommand{\bea}{\begin{eqnarray}}
\newcommand{\eea}{\end{eqnarray}}

\newcommand{\hf}{\frac{1}{2}}
\newcommand{\nn}{\nonumber\\}

\usepackage{graphicx}

\begin{document}
                                                                                                                             
\begin{center}
{\large {\bf Wilsonian renormalization group properties of a novel non-critical cosmological string configuration}} \\
\vspace*{1cm}
{\bf Jean Alexandre} and {\bf Nikolaos E. Mavromatos} \\
\vspace{0.3cm}
Department of Physics, King's College London, London  WC2R 2LS, England \\

\vspace{1cm}
           
{\bf Abstract}

\end{center}

We show that an exact, non-critical-string cosmological configuration
of dilaton and graviton backgrounds, found in \cite{JHEP},
constitutes an infrared stable fixed point of an exact Wilsonian renormalization group equation.

\vspace{0.5cm}

{\it Talk presented at the University of Southampton,
Department of Physics and Astronomy, September 2005.}

\vspace{2cm}

A new non-perturbative time-dependent configuration of the bosonic string was found in \cite{JHEP},
described by the action
\be\label{config}
S=\frac{1}{4\pi\alpha^{'}}\int d^2\xi\sqrt{\gamma}\left\{
\gamma^{ab}\frac{\kappa\eta_{\mu\nu}}{(X^0)^2}\partial_a X^\mu\partial_b X^\nu
+\alpha^{'}R^{(2)}\phi_0\ln(X^0)\right\},
\ee
where $\kappa$ and $\phi_0$ are constants, and it has been shown that this 
configuration is not changed after quantization: it is exactly marginal with respect to $\alpha^{'}$-flows. 
Concerning Weyl invariance conditions, it has been conjectured that these are satisfied in a non-perturbative way
by the configuration (\ref{config}), which is valid in any target space dimension $D$, 
and for any dilaton amplitude $\phi_0$.
The only parameter which is not explicitely known, $\kappa$, depends on $D$ and $\phi_0$, but has 
no cosmological relevance. The corresponding spatially flat Robertson Walker Universe 
has a power-law expanding metric, with a vanishing power when $D-2+2\phi_0=0$. 

\vspace{0.5cm}

We exhibit here the Wilsonian properties of the configuration (\ref{config}),
using the exact renormalization method of~\cite{WH}.
We consider an initial bare theory defined on the world sheet of the string, with cut-off $\Lambda$. The
effective theory defined by the action $S_k$ at the scale $k$ is derived by integrating the ultraviolet
degrees of freedom from $\Lambda$ to $k$. The idea of exact renormalization methods is to
perform this integration infinitesimally, from $k$ to $k-\delta k$,
which leads to an exact evolution equation
for $S_k$ in the limit $\delta k<<k$. The procedure was detailed in~\cite{WH}, and here we reproduce
the main steps. Note that we consider here a sharp cut-off,
which is possible only if we consider
the evolution of the dilaton, as explained now.
                                                                                                                             
We consider a Euclidean world-sheet metric, and we assume that,
for each value of the energy scale $k$, the Euclidean action $S_k$ has the form
\be\label{Euclansatz}
S_k=\frac{1}{4\pi\alpha^{'}}\int d^2\xi\sqrt\gamma
\Big\{\gamma^{ab}\kappa_k(X^0)\delta_{\mu\nu}\partial_a X^\mu\partial_b X^\nu
+\alpha^{'}R^{(2)}\phi_k(X^0)\Big\},
\ee
where $R^{(2)}$ is the curvature scalar of the spherical world sheet.
The integration of the ultraviolet degrees of freedom is implemented in the following way. We write the
dynamical fields $X^\mu$ as $X^\mu=x^\mu+y^\mu$, where the $x^\mu$ are the infrared fields with
non-vanishing Fourier components
for $|p|\le k-\delta k$, and the $y^\mu$ are the degrees of freedom to be integrated out, with
non-vanishing Fourier components for $k-\delta k<|p|\le k$ only.
An infinitesimal step of the renormalization group transformation
reads (we take the limit of a flat world-sheet metric, keeping $R^{(2)}$ finite):
\bea\label{transfo}
&&\exp\left(-S_{k-\delta k}[x]+S_k[x]\right)\\
&=&\exp\left(S_k[x]\right)\int {\cal D}[y]\exp\left(-S_k[x+y]\right)\nn
&=&\int{\cal D}[y]\exp\left(-\int_k \frac{\delta S_k[x]}{\delta y^\mu(p)}y^\mu(p)
-\hf\int_k\int_k\frac{\delta^2S_k[x]}{\delta y^\mu(p)\delta y^\nu(q)}y^\mu(p)y^\nu(q)\right),\nn
&&~~~~~~~~~~~~~~+\mbox{higher orders in}~\delta k, \nonumber
\eea
where $\int_k$ represents the integration over Fourier modes for $k-\delta k<|p|\le k$.
Higher-order terms in the expansion of the action are indeed of higher order in $\delta k$, since each integral
involves a new factor of $\delta k$. The only relevant terms are of first and second order in
$\delta k$ \cite{WH}, which are at most quadratic in the dynamical
variable $y$, and therefore lead to a Gaussian integral. We then have
\bea\label{evolS}
\frac{S_k[x]-S_{k-\delta k}[x]}{\delta k}&=&\frac{\mbox{Tr}_k}{\delta k}\left\{\frac{\delta S_k[x]}{\delta y^\mu(p)}
\left(\frac{\delta^2S_k[x]}{\delta y^\mu(p)\delta y^\nu(q)}\right)^{-1}
\frac{\delta S_k[x]}{\delta y^\nu(q)}\right\}\nn
&&-\frac{\mbox{Tr}_k}{2\delta k}\left\{\ln\left(\frac{\delta^2S_k[x]}{\delta y^\mu(p)\delta y^\nu(q)}\right)\right\}
+{\cal O}(\delta k),
\eea
where the trace Tr$_k$ is to be taken in the shell of thickness $\delta k$, and is therefore
proportional to $\delta k$.
                                                                                                                             
We are interested in the evolution equation of the dilaton, for which it is sufficient to
consider a constant infrared configuration $x^\mu$~\footnote{This is why a sharp cut-off can be used:
the singular terms that could arise from the $\theta$ function, characterizing the sharp cut-off,
are not present, since the derivatives of the infrared field vanish.}.
In this situation, the first term on the right-hand
side of eq.(\ref{evolS}), which
is a tree-level term, does not contribute: $\delta S_k/\delta y^\mu(p)$ is proportional
to $\delta^2(p)$, and thus has no overlap with the domain of integration $|p|=k$.
We are therefore left with the second term, which arises from quantum fluctuations, and
the limit $\delta k\to 0$ gives, with the Ansatz (\ref{Euclansatz}),
\be\label{evolequa}
R^{(2)}\partial_k\phi_k(x^0)=
-k\ln\left(\frac{2\kappa_k(x^0)k^2+\alpha^{'}R^{(2)}\phi^{''}_k(x^0)}
{2\kappa_k(1)k^2+\alpha^{'}R^{(2)}\phi^{''}_k(1)}
\left(\frac{\kappa_k(x^0)}{\kappa_k(1)}\right)^{D-1}\right),
\ee
where a prime denotes a derivative with respect to $X^0$ and
we chose the dilaton to vanish for $x^0=1$.
Eq.(\ref{evolequa}) provides a resummation in $\alpha^{'}$,
since the quantities in the logarithm are the running, dressed quantities. As a result, the
evolution equation (\ref{evolequa}) is exact within the framework of the Ansatz (\ref{Euclansatz}),
and is non-perturbative.
                                                                                                                             
One can easily see that a linear dilaton/flat metric configuration~\cite{ABEN}
\be
\kappa(x^0)=1~~~~~~~~\phi(x^0)=Qx^0,
\ee
where the constant $Q$ is independent of $k$, satisfies the evolution equation (\ref{evolequa}).
This is an exactly marginal configuration, independent of the
Wilsonian scale $k$, which is expected because
it does not generate any quantum fluctuations.
                                                                                                                             
\vspace{0.5cm}
                                                                                                                             
We now come back to the configuration (\ref{config}), and look for a similar exact solution of
the renormalization-group equation (\ref{evolequa}), in which the graviton background is given by
the same expression as in eq.(\ref{config}), but the dilaton has the form
\be
\phi_k(x^0)=\eta_k\ln(x^0),
\ee
where $\eta_k$ is a function of $k$. It is easy to see that such a configuration
indeed satisfies eq.(\ref{evolequa}), provided that $\eta_k$ satisfies $d\eta_k/dk=2Dk/R^{(2)}$,
and hence
\be\label{sol}
\phi_k(x^0)=\left(\phi_0+\frac{Dk^2}{R^{(2)}}\right)\ln(x^0),
\ee
where $\phi_0$ is the constant of integration.
                                                                                                                             
Therefore, we have been able to find an exact solution of the renormalization-group equation (\ref{evolequa}),
which tends to the solution (\ref{config}) in the infrared limit $k\to 0$, with a vanishing derivative
\be
\partial_k\phi_k(x^0)\to 0^{+},
\ee
and thus we conclude that the configuration (\ref{config}) is a Wilsonian infrared-stable fixed point.

\vspace{0.5cm}
                               
We note here that the presence of a logarithmic dilaton (in both the Einstein and string frames),
in a Minkowski Universe, characterizing the solution (\ref{config}), implies
that there will be ``runaway'' coupling constants $g^i$ in the theory, since the latter
are related to the string coupling $e^\phi$. Nevertheless, the rate of change of such couplings
today  are inappreciably small: ${\dot g}^i /g^i \propto 1/t$, where the dot denotes differentiation
with respect to the Einstein cosmic time $t$. It remains to be seen whether in realistic situations
the dilaton develops a potential which could lead to its stabilization.
However, both, the fact that the solution (\ref{config}) is a stable fixed point of a
Renormalization group flow, and the extreme smallness of the changes of the couplings $g^i$ today,
imply that the ``run away'' behaviour of our solution may be sufficient for the correct
phenomenology, without the need for dilaton stabilization.
Further discussion on such issues falls outside our present scope.
                                                                                                                             
We close this note by remarking once again that
the physical implications of the solution (\ref{config}), regarding its r\^ole as a representative of the
exit phase of the linearly-expanding Universe of \cite{ABEN}, have been discussed
in \cite{JHEP}, where we refer the interested reader for further details.

\end{document}